# Liquid dispensing and writing by a nano-grooved pin

Hualai Dong[1,2], Xing Yang[1,2], Cunjing Lv[3], Quanshui Zheng[1,2,4]

**Liquid dispensing and writing in the extremely small size regime are important for applications in many current technologies, such as micro/nano fabrication[1-3], biological/chemical patterning and analysis[4-6], and drug discovery[7,8]. Most of current dispensing/writing methods can be sorted into a category of liquid flowing through tiny tubes or nozzles that requires inputting an impulse energy[9-13], which leads to complex procedures, expensive equipment and narrow material applicability, especially for biomaterials[14]. Here, we report a method that may lead to a new category: liquid flows over the tapered surface of a pin with longitudinal nano grooves on the surface to uninterruptedly perform droplet dispensing and direct writing. The dispensed droplet diameters were controllable from several microns down to 150 nm, and the written line heights were as low as 5 nm. The mechanism underlying automatic liquid storage on conical surface and spontaneous liquid transport through nano grooves is revealed and well modeled by a simple relationship. Furthermore, the nano-grooved pins are much simpler and cheaper in fabrication than nanoscale tubes and nozzles, and pins have much depressed clogging problems that are typically troublesome for tubes and nozzles[13]. Our new strategy may constitute a basis for creating liquid dispensing/writing technologies that are simultaneously smaller, simpler, faster and applicable for more types of materials.**

The key in nano-grooved pin (NGP) technology is that the pin surface contains nano grooves or groove-like nanoscale features. Figure 1d–h shows a typical NGP used in our experiments. This topography image was obtained by scanning electron microscopy (SEM, Quanta FEG 450) and atomic force microscopy (AFM, NT–MDT NTEGRA Prima). The pin was made from a tungsten wire simply by electrochemical corrosion (see Methods). The dispensed liquid in the experiments was a type of epoxy glue (WD3205 A part) or ultraviolet (UV) glue (Lantian 9310) with good fluidity and little volatilization under ambient conditions. The substrate was atomically smooth bare silicon. For direct writing, an additional oxygen plasma treatment was applied to the silicon substrate for 1 min to increase the contact angle hysteresis. After the plasma treatment, the UV glue can maintain the line shape almost without change for ~15 min, which was sufficient for subsequent solidification of the glue lines by UV light exposure.

[1]Department of Precision Instrument, Tsinghua University, Beijing 100084, China; [2]Center for Nano and Micro Mechanics, Tsinghua University, Beijing 100084, China; [3]Institute for Nano- and Microfluidics, Center of Smart Interfaces, Technische Universität Darmstadt, Alarich-Weiss-Straße 10, 64287 Darmstadt, Germany; [4]Department of Engineering Mechanics, Tsinghua University, Beijing 100084, China.

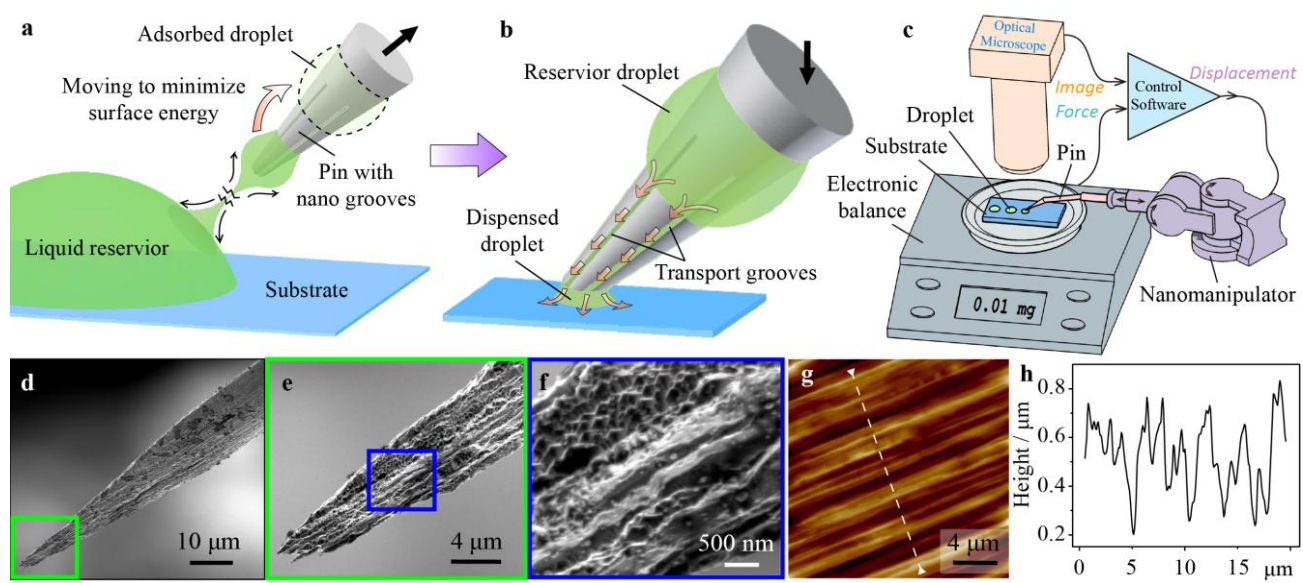

**Figure 1 | Schematics of the liquid adsorbing and dispensing processes by a nano-grooved pin and the pin morphologies. a**, Schematic of the liquid adsorbing process; the adsorbed drop, driven by surface tension force, automatically moves on the conical pin to form a reservoir droplet. **b**, Schematic of the liquid dispensing process through nano grooves onto the substrate spontaneously by capillary force. **c**, Schematic of experimental facilities: the grooved conical pin was installed on a nanomanipulator to perform nanometer-resolution displacement; droplets were dispensed by the pin onto a substrate and observed by an optical microscope; the contact force applied by the pin to the substrate was measured by an electronic balance and further controlled in a closed-loop by computer software. **d–f**, SEM images of a typical nano-grooved pin; **e** is a partially enlarged image of the area marked by the green square box in **d**, and **f** is a partially enlarged image of the area marked by the blue square box in **e**. **g**, AFM image of a typical nano-grooved pin at an area ~3 mm away from the pin end, because this area was relatively more flat and stable than the pin end for AFM imaging. **h**, Cross section in **g** at the position marked by the white dashed line.

Our method mainly consists of two processes: adsorbing liquid and dispensing liquid, as schematically illustrated in Fig. 1a,b. In the first process, an NGP is partly dipped into a large (millimeter-size) liquid reservoir and then pulled out with adsorbed liquid on the pin. The liquid reservoir is a droplet manually transferred by a bamboo stick from the liquid bottle onto the substrate.

The dipping depth is approximately 5–20 μm, and deeper dipping generally leads to more liquid adsorption. With the pin pulled out, it will first suck up a liquid bridge. After the liquid bridge breaks, the liquid adsorbed at the sharp end of the pin will automatically move towards the broad end to form a circular or unilateral droplet, which minimizes surface energy[15,16]. Meanwhile, liquid fills the grooves or groove-like ruggedness on the conical pin to form innumerable nanoscale "fjords", which are analyzed in the discussion section. The second process is dispensing the liquid through the pin onto the substrate (Fig. 1b). The droplet adsorbed on the pin in the first process serves as a reservoir in the second process. When the pin is made contact with the substrate, liquid transports from the reservoir droplet through the nano fjords on the pin to the substrate. After a certain contact time, the pin is lifted, and a small droplet is dispensed on the substrate. This transporting process is spontaneously driven by capillary force. During the contact, the normal force is precisely measured and controlled in a closed-loop. For direct writing, the pin is moved along a designed route while maintaining in contact with the substrate.

In our NGP technology, the aforementioned processes were performed on a vibration isolation platform. As shown in Fig. 1c, the pin was installed on a nanomanipulator (Kleindiek MM3A), and the nanomanipulator was assembled with a 3-axis piezo-positioner (XMT XP–611). The positioner had a 0.1 nm displacement resolution in a 100 μm moving range. The substrate was placed on an electronic balance (Mettler XA205), which served as a force sensor with a 0.1 μN resolution to monitor the normal contact force between the pin and substrate. An optical microscope (Motic PSM–1000) was used to observe the dispensing and writing processes.

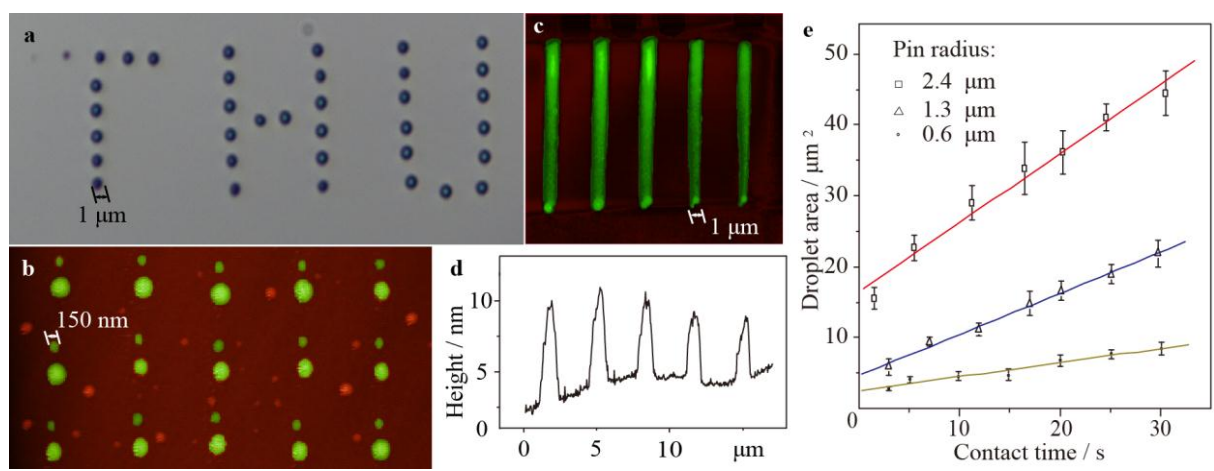

**Figure 2 | Dispensed droplets and written line patterns by nano-grooved pins. a**, Optical image of dispensed droplets with a uniform size by a nano-grooved pin after only one adsorbing process. **b**, AFM image of dispensed nano droplets after solidification. **c**,

AFM image of liquid line patterns consisting of 5 parallel lines. **d**, Cross section of the 5 parallel lines. **e**, The dependence of the dispensed droplet size on the contact time and pin radius; each error bar is the standard deviation for 6 droplets; the three lines indicate linear fits of the discrete points using the least square method. The substrates were polished bare silicon, silicon preprocessed by octadecyltrichlorosilane (OTS), and silicon preprocessed by oxygen plasma in **a**, **b**, **c**, respectively. The liquids used were epoxy glue in **a**,**e**, and UV glue in **b**–**d**. The UV glue was solidified through UV light exposure for AFM imaging, and the glue area is highlighted by green color to distinguish it from the brown silicon substrate for clarity. Each pattern in **a**–**c** was uninterruptedly dispensed or written after only one adsorbing process.

Figure 2a–d shows the typical droplets and line patterns produced by the NGP technology. After only one adsorbing process, an NGP could dispense many uniform droplets without interruption. Maintaining a normal contact force of 10 μN and a contact time of 30 s, the dispensed droplets exhibited excellent size uniformity in diameter of 1 μm. Nano droplets, which had diameters as small as 150–300 nm, were dispensed with a contact force of 0.5 μN and a contact time of 2 s. Each time of contact generated two neighboring droplets in Fig. 2b. Such small droplets exceeded the resolution of ordinary optical microscopy, thus the droplets were imaged by AFM. The contact areas of the droplets on the substrate were approximately linear with the contact time and increased with pin radius (Fig. 2e). The parallel lines (Fig. 2c) were written by an NGP at 1 μm/s, and the line heights were as low as 5–10 nm (Fig. 2d).

In the beginning, pin transfer had been one of the simplest, cheapest and oldest methods, and was widely used for dispensing most liquid materials. Unfortunately, despite these attractive features, pin transfer is being phased out nowadays[17] because it fails to dispense droplets smaller than ~100 μm. In addition, pin transfer is difficult to persistently dispense many droplets with a reasonably good repeatability. The fundamental reason responsible for the above failures was disclosed recently. A driven force, referred to as "*curvi-propulsion*", originates from the curvature gradient of the conical surface, and results in droplets moving spontaneously away from the sharp end of the pin[15,16]. This explanation raises an interesting question: why can an NGP continuously dispense droplets? To reveal the underlying mechanism, we speculated that as the droplet adsorbed by the NGP spontaneously moves from the tip to a broad position on the pin, the nano grooves on the pin surface may capture liquid from the droplet and form innumerable nano "fjords" that are fully or partially filled with liquid, as schematically illustrated in Fig. 3a. The criterion for forming such fjords can be found through considering the free energy, referring to some earlier work[18-20]. The free energy increase per unit of apparent surface area over the fjord region can be calculated as follows: $\Delta u = (f_{LV} - f_{SL} \cos \phi_Y) \gamma_{LV}$, where $f_{LV}$ and $f_{SL}$ respectively denote the relative areas of the liquid–vapor and liquid–solid interfaces over the fjord region per unit of apparent surface area, $\phi_Y$ is the Young's contact angle, and $\gamma_{LV}$ is the liquid surface tension. Thus, the fjord-like wetting is energetically preferred when $\Delta u < 0$. As long as the liquid infiltrates the surface ($\phi_Y < 90°$ and $\cos \phi_Y > 0$), the fjord-type wetting is realizable by introducing a proper ruggedness as follows:

$$\eta \equiv \frac{f_{SL}}{f_{LV}} > \frac{1}{\cos \phi_Y} \tag{1}$$

For examples, the values of $\eta$ for the three simply shaped grooves as illustrated in Fig. 3b are $\eta = 1 + \frac{2h}{w}$, $\sqrt{1 + \left(\frac{2h}{w}\right)^2}$ and $1 + \frac{\pi h}{w}$, respectively. Thus criterion (1) can be easily satisfied through increasing the groove depth $h$ or decreasing the groove width $w$.

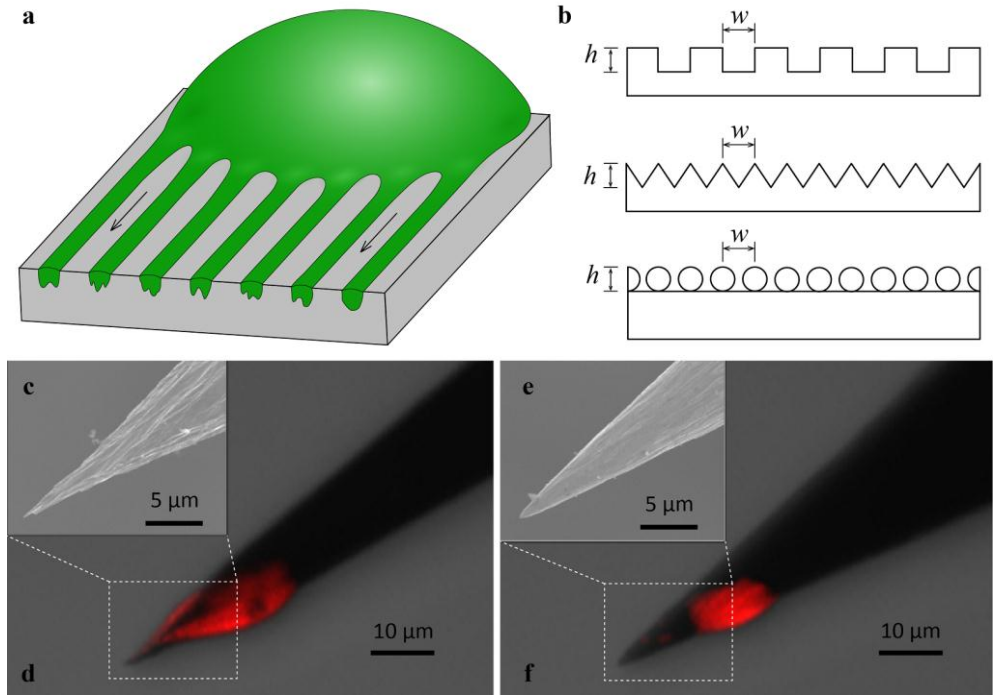

**Figure 3 | Wetting model and experimental confirmation of liquid distribution on pins with different ruggedness. a,** Wetting model, where liquid flows from the droplet into the grooves forming into fjords. **b,** Parameter definitions of different cross sections in the wetting model. **c,** SEM image of a grooved conical pin. **d,** liquid spreading continuously from the reservoir to the pin end. **e,** SEM

image of a smooth conical pin without obvious grooves. **f**, Fluorescent image of liquid forming into a droplet far from the pin end. The fluorescent images were acquired by LSCM and overlaid with bright-field images. The liquid was epoxy glue mixed with rhodamine B to produce fluorescence under excitation by a 543-nm laser.

To confirm that this wetting state is the mechanism of NGPs' capability of persistently dispensing and writing, fluorescent images were taken by laser scanning confocal microscopy (LSCM, Zeiss 710) to observe the liquid distribution on different pins. The fluorescence was produced by epoxy glue mixed with rhodamine B under excitation by 543-nm laser. Figure 3c,d shows that liquid fjords spread over the entire region from the reservoir droplet to the pin end. In contrast, an ordinary pin whose surface was relatively smooth did not have continuous liquid channels, as indicated in Fig. 3e,f, and thus was incapable to persistently dispense liquid.

The above understanding can guide the design of NGPs for liquid dispensing or writing in the resolution of tens of nanometers, as there exist technologies for fabricating surface roughness in a few nanometers such as carbon nanotube bundles[21]. Besides, our other experiments showed that conical surfaces with grooves in micrometer or even millimeter scale could still support liquid flowing over the grooves and allow liquid dispensing. Therefore, the NGP technology is expected to be valid at the scale ranging from tens of nanometers to several millimeters.

Next, we must specially discuss another technology, dip-pen nanolithography (DPN), because both NGP and DPN use pins with nanoscale tips. DPN deposits molecules by AFM tips. This technology can reach sub-50-nm resolution[22,23] and is bio-compatible for DNA and cells. However, the material transfer process of DPN is molecule deposition, not liquid dispensing, according to its principle of water meniscus condensed from atmosphere humidity. The mechanism of DPN is completely different from that of our NGP technology. The key in DPN is the condensed water meniscus, whereas in NGP, the key is the grooves. DPN has generally been used to modify surface properties with a few molecular layers, rather than directly used for additive manufacturing when thickness more than tens of nanometers is needed[24]. Moreover, liquid-state materials and non-water-soluble materials are difficult to apply by DPN. Another similar example is scanning force microscopy (STM) lithography[25], which can even reach atomic level resolution but is not suitable for massive micro/nano fabrication.

Finally, Fig. 4 is presented for comparing NGP with other major liquid dispensing technologies in the phase space of producible drop size versus technology complexity. Previously, the main road for liquid dispensing technology development was based on tubes or nozzles from the original pipette with a pump/valve[10] to ink jet[11] and then to electrohydrodynamic (EHD) jet[12,13,26,27], as indicated by the grey arrows. The size of dispensed droplets gets smaller and the dispensing speed gets faster but at the expense of the methods being very complex and only applicable for quite limited types of materials. Now we find another road that may revive the old pin transfer technology. Compared to the existing liquid dispensing methods, NGP technology has the advantages of simplicity, low cost, anti-clogging, and size controllability in micro/nano scales, and it is also promising for application in a variety of fields including micro/nano fabrication, biological/chemical patterning, drug discovery and lab on a chip. In the future, the combination of NGP and EHD may accelerate the liquid dispensing speed and expand material applicability.

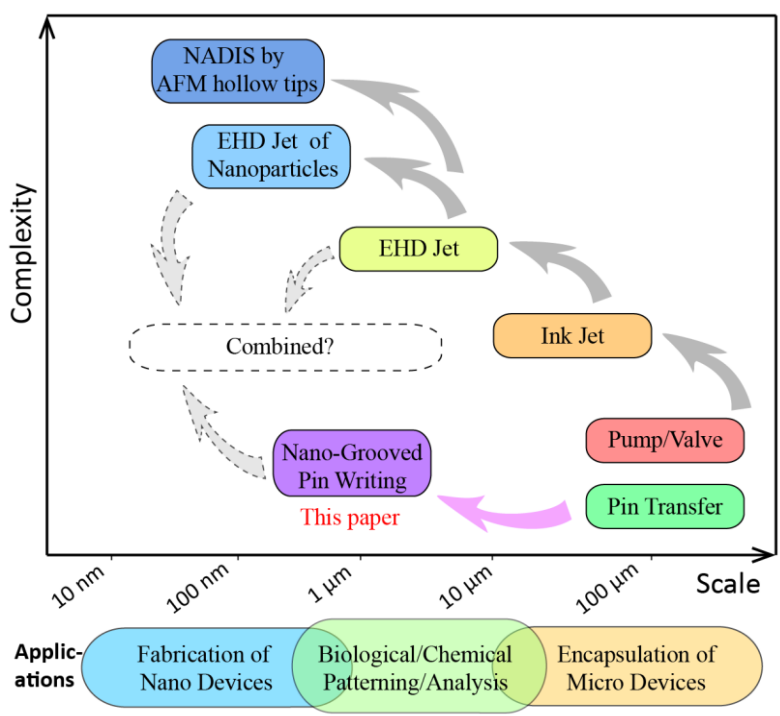

**Figure 4 | Roadmaps of major liquid dispensing technologies.** Traditional routes of liquid dispensing are based on tubes or nozzles developing from pump/valve to ink jet and then to EHD jet. Although the size gets smaller and the speed gets faster, the complexity increases dramatically, and certain material requirements to the liquid and substrate must be satisfied. To further downscale, EHD jet of nanoparticles was used in nano fabrication, and AFM hollow tips with nano apertures were used for nano dispensing (NADIS)[12,24,28]. Apart from very complex hollow fabrication, dispensing speed was slowed down by flow resistance in nanoscale and the concomitant clogging problem was troublesome. We present a new route based on nano-grooved pins that downscales the pin transfer while

maintaining similar simplicity and material applicability. In the future, EHD jet may be simplified, and our NGP technology has the potential to speed up by parallel strategy[29,30] and the potential to be further downscaled by sharper pins. The best technology for liquid dispensing in the future may be some method that combines the two strategies and requires either electrical conditions or wetting conditions for the materials.

## Methods

The nano-grooved pins were fabricated by alternating current (AC) electrochemical etching. A tungsten wire of ~0.3 mm in diameter, ~25 mm in length and purity over 99.95% was dipped into ~40 mL of 12 mol/L KOH aqueous solution in a 50 mL beaker. The dip depth of the tungsten wire was ~8 mm, and a circular copper electrode surrounding the dipped tungsten wire was also dipped into the KOH aqueous solution. An alternating current of 50 Hz was applied between the tungsten wire and the circular copper electrode by an adjustable AC transformer. The electrochemical etching included four main steps. Step 1, the applied electric voltage was set to amplitude of ~55 V until the dip depth of the tungsten wire decreased to ~2 mm from ~8 mm as the electrochemical etching progressed. Step 2, the applied electric current was turned off, and the etched tungsten wire was immersed in deionized water with ultrasonic cleaning for 3 s. Note that the sharp end of the etched tungsten wire should not come into contact with anything other than water or air. Step 3, the washed tungsten wire was dipped into the KOH aqueous solution again with a dip depth of ~0.5 mm. The applied electric voltage amplitude was set to ~8 V until the dipped segment was etched out. Step 4, the twice-etched tungsten wire was removed and immersed in deionized water for 3 s. A sharp pin was successfully made from the tungsten wire. Note that ultrasonic cleaning should not be applied to the twice-etched tungsten wire as severe vibration may damage the sharp pin. Usually, a batch of such pins were fabricated and observed by SEM, but only a small portion of the pins had obvious continuous grooves. These grooved pins were chosen for dispensing droplets or direct writing. The mechanism of formation of the nano grooves is still unclear, and thus, the surface topography of the pins differs greatly. For the pins with relatively smooth surfaces in Fig. 3e, the fabrication procedure was the same as described above, except that KOH aqueous solution of 4 mol/L was used.

**Acknowledgements** We thanks for the financial support from NSFC (No. 51375263) and National Key Basic Research Program of China (No. 2013CB934200).

**Author Contributions** H.D. and X.Y. found the phenomenon and analyzed the first data; H.D. performed the experiments; Q.Z. found the mechanism and gave together with C.L. the theoretical explanation; H.D. and Q.Z. co-wrote the manuscript; Q.Z. conducted the research.

**Author Information** The authors declare no competing financial interests. Correspondence and requests for materials should be addressed to Xing Yang (yangxing@tsinghua.edu.cn) and Quanshui Zheng (zhengqs@tsinghua.edu.cn).

# Supplementary Information

# Liquid dispensing and writing by a nano-grooved pin

Hualai Dong, Xing Yang, Cunjing Lv, Quanshui Zheng

This file includes:

9 sections (Movie descriptions, Materials, Adsorbing and dispensing process, Droplet size dependence, Contact angle measurement, Wetting on grooved surface, Dispensing liquid in millimeter scale, Image acquisition and processing, Comparable technologies);

8 figures (Figs S1–S8);

3 tables (Tables S1–S3);

References.

## Movie descriptions

Movie S1: The process of dispensing the droplet pattern in Fig. 2a. The playback speed was 90 times the normal speed. The diameters of these uniformed droplets were 1 μm.

Movie S2: The process of directly writing the line pattern in Fig. S1b. The playback speed was 12 times the normal speed. The line width of the pattern was 5μm. Figure S1b shows the optical image of the parallel lines whose AFM image has been shown in Fig. 2c (shown again in Fig. S1a). Distinguishing the lines from the silicon substrate in the optical image is difficult because the line height is only 5–10 nm and the line width is only ~1 μm, which is near the resolution of optical microscopy. For this reason, the shown movie of the directly writing process was chosen as the 30-μm-side square instead of the 5 parallel lines.

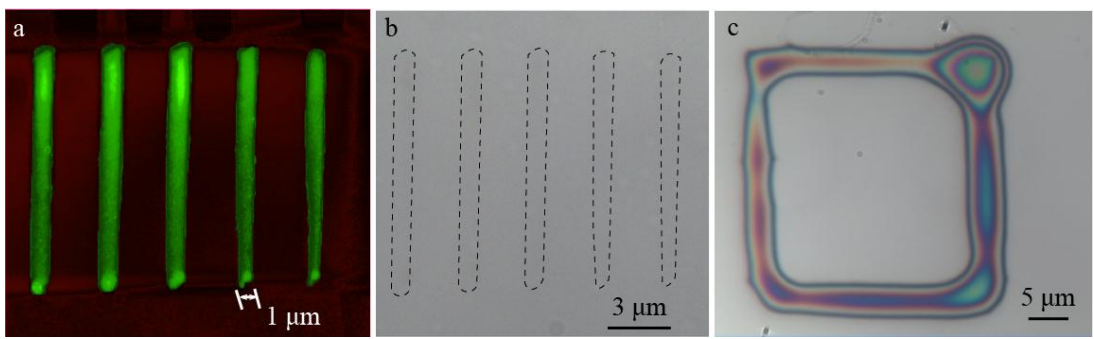

**Figure S1. Liquid line patterns written by grooved pins. a**, AFM image of 5 parallel lines shown in Fig. 2c. **b**, Optical image of the same 5 parallel lines which were marked by dashed lines. **c**, Liquid line pattern of a 30-μm-side square.

## Materials

Two kinds of liquids are used to perform the experiments. The first kind of liquid material is epoxy glue, which is the A part of commercial AB glue (WD3205). The AB glue includes two parts: A part and B part. These parts will remain in their liquid states unless mixed together. Only the A part was used in our experiments to prevent this part from drying or solidifying. The second kind of liquid material is commercial UV glue (Lantian 9310), which will remain in its liquid state unless exposed to UV light. These kinds of liquid materials have good fluidity before they reach their solid states by mixing or exposure. The epoxy glue was used to dispense microscale droplet pattern (Fig. 2a) and to test the size dependence of contact time and the radius of pins (Fig. 2e). The UV glue was used to dispense nanoscale droplets (Fig. 2b), which could be solidified by UV exposure for measurement by AFM. A UV light source with a characteristic wavelength of 325 nm and power of 112 W was shone on the nanoscale droplets for 2 min to solidify the UV glue. For direct writing, the UV glue was chosen because of its relatively larger flow hysteresis, thus easier to keep the line shape instead of shrinking back into droplet shape (Fig. 2c). Epoxy glue mixed with rhodamine B was used on the LSCM images (Fig. 3d,f) to produce fluorescence for observing liquid distribution on conical pins.

Bare silicon wafers were polished to near atomic smoothness and were used as the substrate for dispensing droplets and direct writing. The wafers had an initial diameter of 4-inch and were then cut into 10 mm × 8 mm rectangle slices with protection by blue films. The square slices were washed by ultrasonic cleaning in ethanol for 16 min. The topography of a silicon slice was measured by AFM to confirm its near-atomically smooth surface. Dispensing or writing process was performed within 12 h after the washing procedure to avoid excessive dust pollution. This kind of bare silicon was used to dispense microscale droplets (Fig. 2a). Additional surface treatments of OTS and oxygen plasma were applied to the washed bare silicon slices for nano-droplets and direct writing, respectively.

For dispensing nano droplets, OTS treatment of bare silicon substrate aimed to increase the liquid-substrate contact angle and thus increase the droplet height for the convenience of AFM imaging (Fig. 2b). The bare silicon slices were washed in acetone by ultrasonic cleaning for 10 min and then dried by nitrogen. In the following procedures, two kinds of solutions were used to soak the silicon slices. Piranha solution was compounded by $H_2SO_4$ and $H_2O_2$ at a volume ratio of 7:3. OTS solution was compounded by OTS and n-Hexadecane at a volume ratio of 250:1. First, the silicon slices were immersed in piranha solution in 90 ℃ water bath for 30 min. Second, the silicon slices were washed by deionized water and then dried by nitrogen. Third, the silicon slices were immersed into OTS solution in 45 ℃ drying oven for 20 min. Fourth, the silicon slices were washed by through immersion in $CHCl_3$ for 15 min and then in ethanol for 30 min. Finally, the silicon slices were collected and dried by nitrogen. All the reagents used in the above procedures were analytical reagents.

For direct writing, oxygen plasma treatment was applied to the bare silicon substrate to increase the hysteresis of liquid on the substrate (Fig. 2c). The oxgen plasma treatment lasted for 1 min under a chamber pressure of 0.5 mbar (50 Pa) and 30 W power (Femto PCCE) after the washing procedure and before the writing procedure. Considering the limited efficient time of 1 min plasma treatment, the writing procedure was done within ~3 h.

## Adsorbing and dispensing process

The main procedures of pin fabrication were introduced in the Methods part. The fabrication method is simple and low cost. However, the shape and topography of the electrochemically etched pins are difficult to control efficiently as desired. For this reason, the screening procedure is very important to obtain pins with obvious continuous grooves. Using images by SEM or AFM can help with screening to some extent but is inadequate and inefficient. The practical way to judge a pin is through direct testing of the continuously dispensing capability of the pin. Relatively smooth pins were generally incapable of dispensing any droplets. However, if a pin can dispense liquid when in contact with substrate, the pin does not always have continuous grooves. Two possible mechanisms may be responsible: pin transfer and nano-grooved pin writing. Continuity is the key to distinguish the two mechanisms. If the pin transfer mechanism is working, then the pin can dispense only one or several droplets. In addition, the size of the droplet decreases rapidly as the dispensing progresses after one-time adsorbing process. Moreover, if the pin adsorbs more liquid, then it may have difficulty dispensing any droplets because a larger reservoir droplet will move to a farther position from the pin end according to curvi-propulsion theory[15,16]. As shown in Fig. 3f, the reservoir droplet stayed far from the pin end, and thus, liquid has difficulty transferring to the substrate when the pin was contacted to the substrate. By contrast, if the mechanism of nano-grooved pin writing is working, then the pin can dispense many droplets continuously after only one-time adsorption. The size of the droplet decreases slightly as the dispensing number increases. In our experiments, some pins can continuously dispense more than one hundred droplets of similar size. In this case, larger adsorbed reservoir droplet on the pins will facilitate better continuity.

The adsorbing process is shown in Fig. S2. In usual experiments, the illumination of the optical microscope was in reflection mode and the droplets were observed from the top view. To observe the adsorbing process clearly, we used illumination in transmission mode and observed the droplet reservoir from the side view. The droplet adsorbed by the pin in Fig. S2f appeared to be circular on the conical pin. However, in usual experiments, the adsorbing process may tend to obtain unilateral droplets. When observed from the top view by a microscope, the pin was pulled out in the upward direction rather than the direction parallel to the pin axis. The pin axis was neither perpendicular nor parallel to the substrate and has an incline angle of 30 °–60 °. As a result, both circular and unilateral droplets could be adsorbed on the conical pin. Moreover, in both cases, the droplets will move away from the pin end to minimize total surface energy.

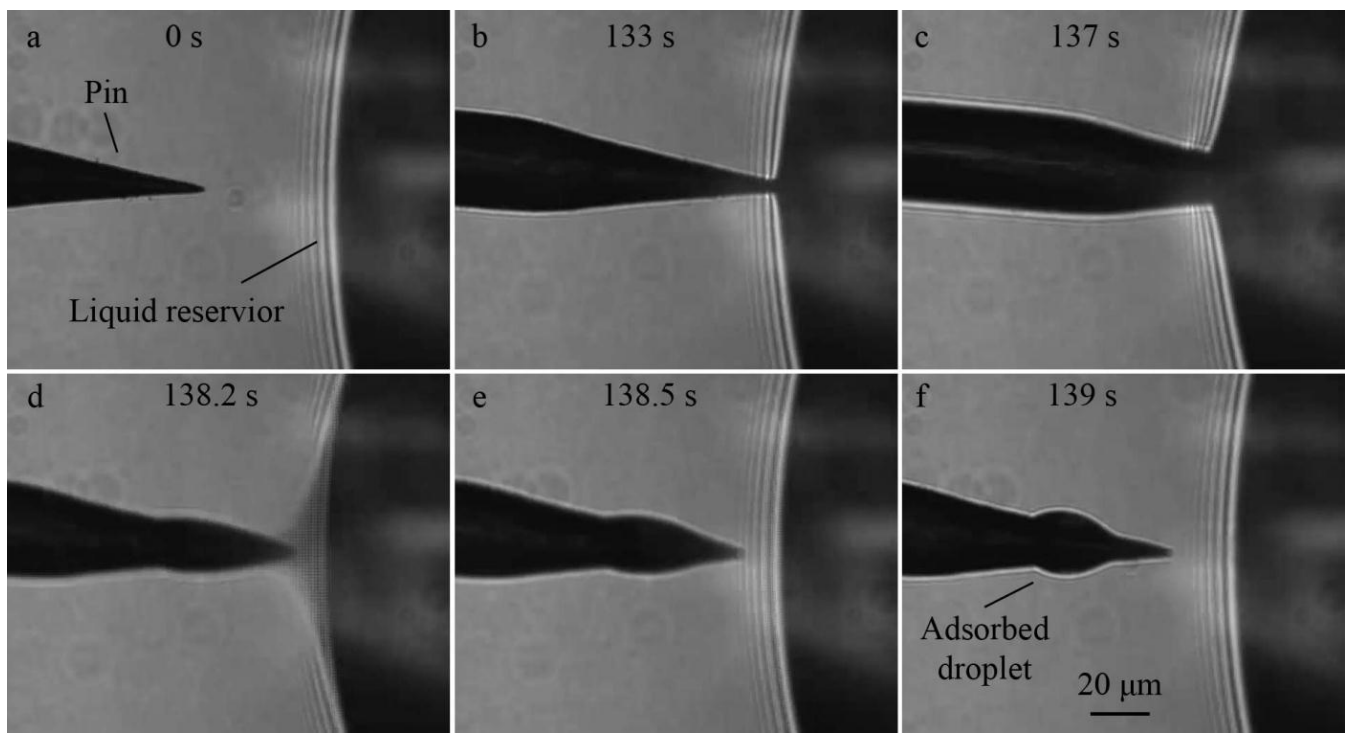

**Figure S2. Procedures of adsorbing liquid by a pin. a**, The pin approached the liquid reservoir on the substrate. **b**, The pin end touched the surface of the liquid reservoir. **c**, The pin was further dipped into the liquid reservoir and sucked up a liquid bridge. **d**, The pin was pulled out and the liquid bridge was breaking. **e**, The pin adsorbed some liquid; the liquid shrank into a droplet, and the droplet moved away from the pin end. **f**, A droplet was formed on the conical pin, and this droplet could serve as a liquid reservoir in the following dispensing process. The images were taken by an optical microscope (Olympus BX51) in transillumination mode and from the side view. Gray processing was applied to the images that originally had distorted colors. The scale bars in **a–e** were the same as the scale bar in **f**.

In addition, some cases exist in which some pins could dispense liquid continuously after adsorbing a small amount of liquid, but lost this capability when adsorbing a large amount of liquid. We consider the reason probably to be that the nano grooves did not spread continuously enough to form enough long fjords or the ruggedness of the grooves is insufficient all the way from the reservoir droplet to the pin end. In practice, such cases are quite common whereas good dispensing continuity is infrequent. Consequently, the adsorbing process is very import to examine the dispensing capability of a pin. To avoid excessive adsorption to a pin, we first dip the pin at a shallow depth into the reservoir droplet on substrate to adsorb a small amount of liquid. The pin was moved slowly by a nanomanipulator under the observation of an optical microscope. Once the pin made contact with the liquid reservoir, it was pulled out by the nanomanipulator as quickly as possible. In this way, only a small amount of liquid was adsorbed onto the conical pin such that the adsorbed droplet would not move too far from the pin end.

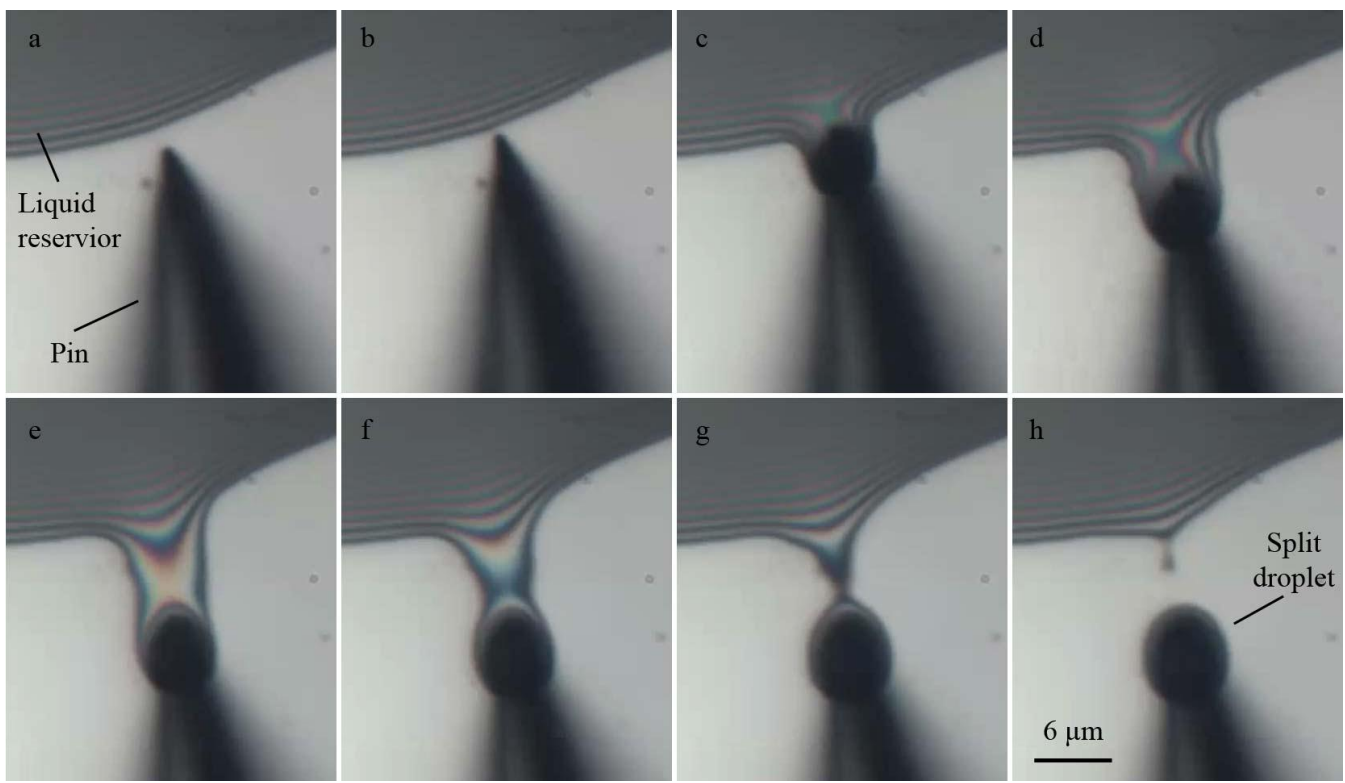

**Figure S3. Procedures of splitting the reservoir droplet on the substrate by a pin. a–b,** The pin approached the liquid reservoir while maintaining contact with the substrate. **c,** The pin touched the liquid reservoir and adsorbed a liquid bridge. **d–e,** The pin was moved away while still maintaining contact with the substrate. **f–g,** The liquid bridge broke because of surface tension force. **h,** A small droplet was successfully split from the large liquid reservoir on the substrate, and this droplet could act as a small liquid reservoir in the following adsorbing process. The liquid material was epoxy glue, and the substrate was polished bare silicon after oxygen plasma treatment. The images were taken by an optical microscope (PSM-1000) from the top view. The scale bars in **a–g** were the same as the scale bar in **h**.

Another simpler method (Fig. S3) was also used to ensure that the pin adsorbs only a small amount of liquid. The pin was moved to dip into the liquid reservoir on the substrate at the edge of the liquid reservoir and then moved down to contact with the substrate. Next, the pin was moved away from the liquid reservoir while maintaining contact with the substrate and then split out a small liquid reservoir with a diameter of 5–50 μm. Finally, the pin was lifted from the small liquid reservoir to adsorb only a small amount of liquid. If the pin was capable of dispensing continuously after such a small amount of adsorption, a larger amount of adsorption could be considered next time. If a pin adsorbed an excessive amount of liquid, then it could be recovered by washing in organic solvent, such as ethanol or acetone. Notably, washing with ultrasonic vibration for a long time could damage the shape and topography of the pin. In this asymptotic way, many nano-grooved pins could work with a certain continuity even when their grooves were not fully continuous.

The dispensing process was usually observed under an optical microscope to confirm the contact between the pin and the substrate, and to monitor the dispensed droplets or lines in real time. An electronic balance was used as a force sensor to confirm the contact when the pin approached the substrate and to monitor the normal force applied by the pin to the substrate. The electronic balance has a resolution of 0.1 μN. Thus, a wind shelter made of transparent plastic was used to protect the balance pan from air turbulence. To ensure that the plastic produced no electrostatic interference to the measured force, two pre-measurement force values were determined before and after contacting the pin to the substrate. If the two force values had a difference no more than 0.3 μN, then we could confirm that there was little or no electrostatic interference. The contact force applied in our experiments varied from 0.5 μN–50 μN, depending on the desired droplet size or line width and pin stiffness. A larger force was usually applied when larger droplets were desired or the pin had a higher stiffness. However, the contact force has an insignificant influence on the droplet size or the line width. The two parameters depend more on the contact time and pin radius.

Direct writing is more complex than dispensing droplets, because this process requires the liquid material to have enough hysteresis to prevent line shrinking force. For a droplet on substrate, the contact angle hysteresis is defined as the difference between the advancing contact angle and the receding contact angle. If there was no contact angle hysteresis, a static small droplet would be a sphere cap because this shape gives the minimal surface energy. The contact angle hysteresis allow a droplet to have different local contact angles at different boundary positions, possibly leading to the shape of a line instead of a sphere cap. Apart from the contact angle hysteresis, the liquid viscosity also plays an important role in dynamic situations. Although the line shape tends to shrink back into droplet shape, overcoming the liquid viscous resistance may take a long time for the surface tension force. In direct writing experiments, the liquid line could keep its shape for more than 10 min which was adequate for the subsequent solidification procedure by UV light.

Given that the dispensing results were influenced by the material properties, it is worth to note that the UV glue will gradually become increasingly viscous under natural light. In theory, the properties of UV glue will not change until UV light exposure according to user instructions. However, we think that the natural light and the illumination of the optical microscope also contain a small amount UV light, which may gradually influence the viscosity and hysteresis of UV glue. Exposure to air may also contribute to the gradual change of viscosity and hysteresis. In practice, the UV glue was observed having a noticeable change in viscosity (and possible in contact angle hysteresis) within several hours after being taken out from the glue bottle. Moreover, we have noted that the UV glue taken from the surface layer of the remaining liquid in the glue bottle tends to have larger viscosity than the glue taken from the inner part. Suitable liquid viscosity is necessary because the viscosity needs to be high enough to keep the line shape for a long time, but in the same time the viscosity also needs to be low enough to obtain sufficiently good fluidity for continuously transporting. The gradual change in viscosity and hysteresis actually benefits our direct writing experiments. In particular, we have chance to find the most suitable viscosity and hysteresis only through repeated attepmpts every half an hour. The best time may be 1−5 hours after the UV glue was taken out from the glue bottle. When the UV glue was used to dispense droplets, the viscosity should be as low as possible. In this case, the UV glue was taken from the inner part of the remaining liquid in the glue bottle, and the following procedures were performed within 2 h. With regard to the epoxy glue, we observed no change in viscous or hysteresis properties even several days after it was taken out.

## Droplet size dependence

Varying the contact time and the pin radius could effectively control the size of the dispensed droplets, as shown in Fig. 2e. Each relationship line in Fig. 2e was a linear fit of 7 discrete points, and each point was the average contact area of 6 droplets. The error bars were the standard deviation calculated from the 6 droplets. A total of 126 droplets were used for the relationship lines. These droplets were shown in Fig. S4. Numerical data of the 21 discrete points were presented in Table S1.

It seems that the droplet contact area is approximately linear with the contact time for one single pin. However, this conclusion should be given serious consideration. The change of the droplet size was not very big, approximately two or three times for one single pin. Moreover, the initial stages of the fitted lines did not approach zero when the contact time was near zero. Given the limited stored liquid on the conical pin, the maximal droplet size was also limited. The dispensing speed also tended to slow down as the stored liquid was exhausted after dispensing many droplets. A theoretical explanation of size dependence was not provided in this paper, because this issue is quite complex and may involve fluid dynamics. However, a thorough study of the dynamic behavior in the liquid dispensing process that involves liquid transporting through nano fjords would be a potential research direction. In this sense, our experiments may be used to study or verify the physics in nanofluidics[31].

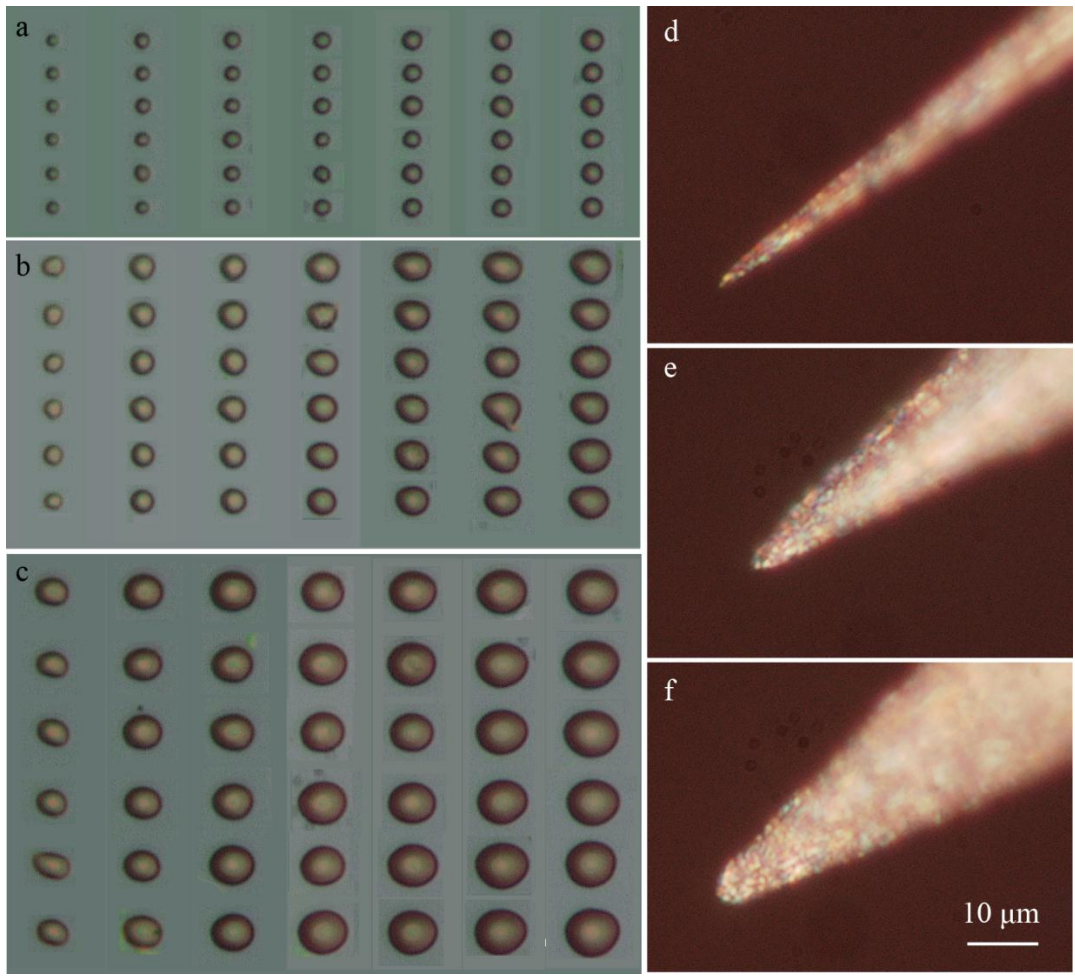

**Figure S4. Differently sized droplet arrays dispensed by three pins with different pin radius. a,b,c** were the droplet arrays dispensed by the pins **d,e,f**, respectively. All the images were taken by an optical microscope (Hirox KH-3000) in reflection illumination mode. The liquid material was epoxy glue, and the substrate was polished silicon with a 330 nm oxide layer. Postprocessing was applied to the images of the droplet arrays such that the droplets were moved to align in grids. The scale bars in **a–e** were the same as the scale bar in **f**.

**Table S1. Numerical data of the droplets in the size dependence relationship**

| Pin radius | Value name | Value unit | Values | | | | | | |
|---|---|---|---|---|---|---|---|---|---|
| 0.6 μm | Contact time | s | 2.97 | 5.05 | 9.87 | 14.9 | 20.1 | 25.1 | 30.1 |
| | Average contact area of droplets | μm$^2$ | 2.80 | 4.07 | 4.53 | 4.72 | 6.77 | 7.69 | 8.40 |
| | Standard deviation of contact area | μm$^2$ | 0.224 | 0.492 | 0.603 | 0.801 | 0.754 | 0.671 | 0.891 |
| 1.3 μm | Contact time | s | 2.93 | 6.98 | 11.9 | 17.0 | 20.1 | 25.1 | 29.8 |
| | Average contact area of droplets | μm$^2$ | 5.86 | 9.35 | 11.1 | 14.8 | 16.6 | 18.9 | 21.9 |
| | Standard deviation of contact area | μm$^2$ | 1.17 | 0.632 | 0.846 | 1.68 | 1.33 | 1.33 | 1.87 |
| 2.4 μm | Contact time | s | 1.52 | 5.53 | 11.3 | 16.5 | 20.3 | 24.6 | 30.5 |
| | Average contact area of droplets | μm$^2$ | 15.5 | 22.7 | 29.0 | 33.9 | 36.1 | 41.0 | 44.5 |
| | Standard deviation of contact area | μm$^2$ | 1.49 | 1.76 | 2.42 | 3.66 | 3.04 | 1.85 | 3.17 |

## Contact angle measurement

The wetting condition for dispensing liquid by a nano-grooved pin is given by Formula (1). The contact angles of liquid on solid materials are important to satisfy the formula. Accordingly, these contact angles were measured. In addition, the topography of the grooved pin was measured, and the ruggedness of the fjords was analyzed. Then, the contact angle and the fjord ruggedness were combined to verify that Formula (1) was successfully satisfied.

First, the Young contact angles of used epoxy glue and UV glue on the tungsten material were measured by a contact angle meter, as shown in Fig. S5. Directly measuring the contact angle on the conical pin is difficult. Thus, a tungsten (the material of the pins) block with a mechanically polished surface was used as the substrate to measure the contact angle. The tungsten block was ultrasonically washed in ethanol for 16 min. A droplet with a size of 1–2 mm was placed on the tungsten surface. This droplet was observed to be in an anisotropic wetting state because the polished tungsten surface had obvious textures. The contact angles measured from the direction parallel to the texture and the direction perpendicular to the textures were remarkably different. Previous studies have investigated the influence of grooves on the contact anlges[32]. The contact angle measured from the parallel direction was significantly influenced by the surface grooves because of the increased contact angle hysteresis. However, the contact angle measured from the perpendicular direction was nearly the same as that on the flat surface[32]. Thus, we took the contact angles measured from the perpendicular direction as the Young contact angles, which will be used in the following analysis to verify Formula (1).

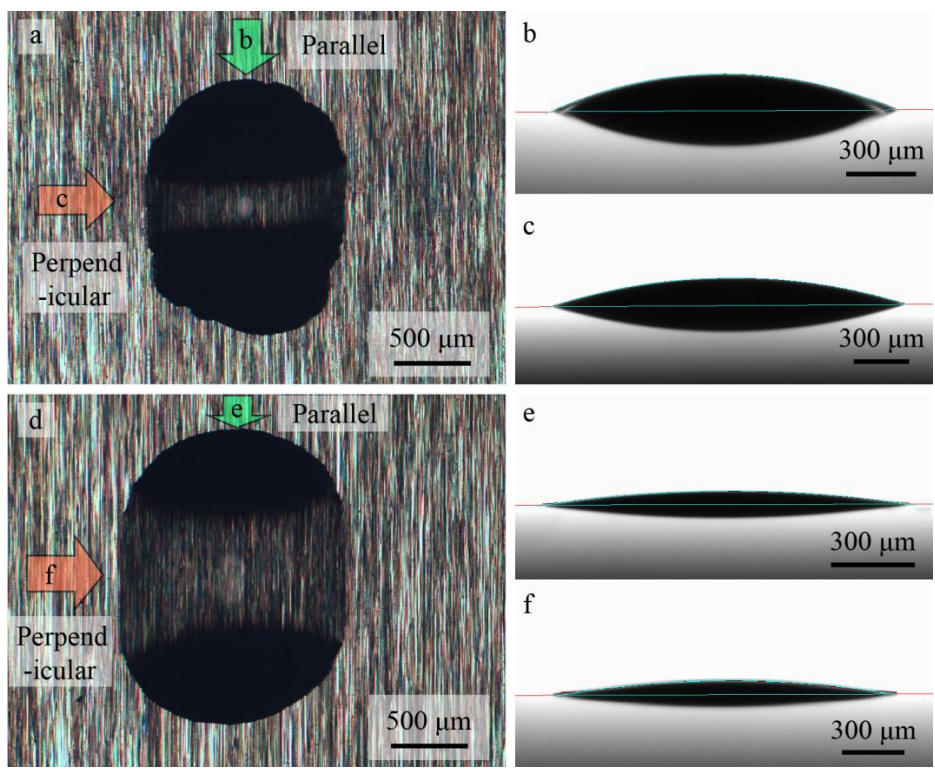

**Figure S5. Droplets for contact angle measurement. a**, Epoxy glue droplet on tungsten block in top view. **b**, Side view image of the epoxy glue droplet in **a** observed from the direction parallel to the texture direction of the tungsten surface. **c**, Side view image of the epoxy glue droplet in **a** observed from the direction perpendicular to the texture direction of the tungsten surface. **d**, UV glue droplet on tungsten block in top view. **e**, Side view image of the UV glue droplet in **d** observed from the direction parallel to the texture direction of the tungsten surface. **f**, Side view image of the epoxy glue droplet in **d** observed from the direction perpendicular to the texture direction of the tungsten surface. The textures on the tungsten surface were produced by mechanical polishing. The color images **a,d** were taken by an optical microscope (PSM-1000), and the gray images **b,c,e,f** were taken by a contact angle meter (OCA20). The red base lines and cyan contour lines in **b,c,e,f** were automatically calculated by the data processing software of the contact angle meter. The fitting algorithm was chosen as the Young–Laplace fitting in the data processing software.

The measured contact angles are presented in Table S2. Apart from the contact angles between liquid and tungsten, the contact angles between the liquid and the substrates were also measured. A liquid and a substrate should infiltrate each other when dispensing the liquid to the substrate. In other words, the contact angle of the liquid on this substrate should be smaller than 90 ° because if the contact angle is larger than 90 °, then no liquid will be left on the substrate when the pin was lifted from the substrate without considering hysteresis and viscosity. The data in Table S2 indicates that all the contact angles of the two kinds of glue on the four kinds of substrate (excluding the combinations not used in our experiments) were smaller than 90 °.

**Table S2. Contact angles of different liquids on different substrates**

| Substrate | Contact angles / degree | |
| :---: | :---: | :---: |
| | Epoxy glue | UV glue |
| Tungsten block (observation direction is parallel to texture) | 25.5 ±1.9 | 7.0 ±5.3 |
| Tungsten block (observation direction is perpendicular to texture) | 18.6 ±1.8 | 6.0 ±2.3 |
| Polished bare silicon | 30.4 ±6.8 | / |
| Polished bare silicon after OTS treatment | / | 50.1 ±8.5 |
| Polished silicon with 330 nm oxide layer | 35.9 ±8.5 | / |
| Polished bare silicon after oxygen plasma treatment | / | 10.2 ±1.9 |

Another factor used in Formula (1) was ruggedness of the fjords, which was the ratio of the relative areas of the liquid–vapor and liquid–solid interfaces over the fjord region. To obtain the ruggedness value, we first obtained the topography of a nano-grooved pin used in our dispensing experiments by AFM, as shown in Fig. S6. Given that the topography at the pin end could not been obtained clearly by AFM, we took the topography at an area ~3 mm away from the pin end to check the wetting conditions. Three grooved parts were taken out to analyze the ruggedness and determine if they were capable of forming fjords. These grooved parts are marked by red, green, and blue colors respectively in Fig. S6c.

The ruggedness was calculated by using the actual length of the curve path divided by the straight length between the two endpoints of the curve path. The actual length was calculated by accumulating the straight length of tens of sub-paths defined by the acquired data points. The straight length was defined by $\sqrt{(x_1 - x_2)^2 + (y_1 - y_2)^2}$, where $x_1$, $y_1$ and $x_2$, $y_2$ are the coordinate values of the two endpoints, respectively. The final calculated ruggedness together with the minimal ruggedness to form fjords is shown in Table S3. All the three grooves as well as the whole section have a ruggedness larger than the required minimal ruggedness. Accordingly, they satisfy Formula (1). Notably, satisfying Formula (1) is easy for a liquid like the UV glue, which has a small contact angle on tungsten. Furthermore, even if the contact angle is too large to satisfy the wetting condition, certain surface treatment, such as simple oxygen plasma, can lower the contact angle easily, then the wetting condition can be satisfied. Therefore, both the liquid and the substrate can satisfy the wetting condition of Formula (1) easily.

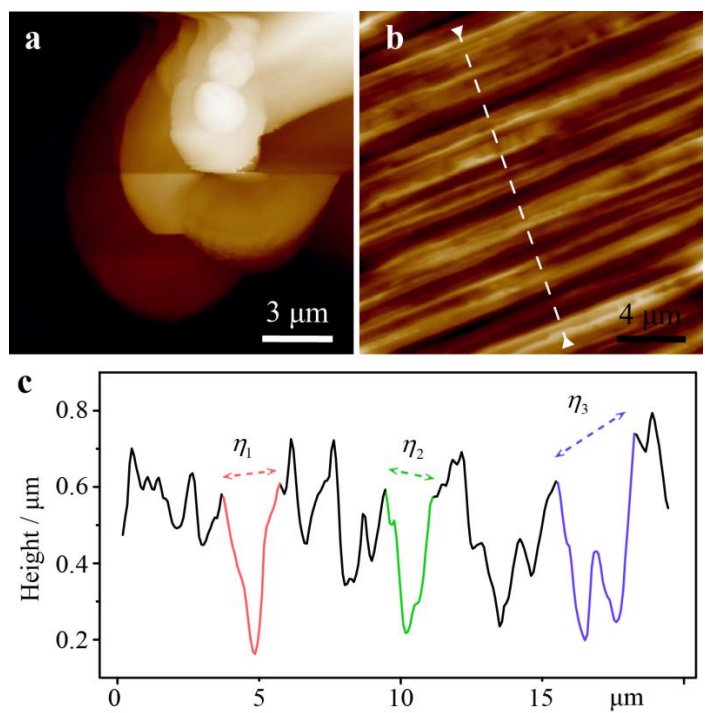

**Figure S6. Topography of a grooved pin. a**, Topography image of a grooved pin at the position of the sharp end; this image is fuzzy because the pin end was too soft to bear the force applied by AFM imaging. **b**, Clear topography image of a grooved pin at an area ~3 mm away from the pin end; this area is relatively more flat and stable than the pin end for AFM imaging. **c**, Cross section in **b** at the position marked by a white dashed line. Three grooves that may form into fjords were highlighted in the curve by red, green, and blue colors. Given that the topography at the pin end could not be obtained clearly by AFM, we used the topography at an area ~3 mm away from the pin end to check the wetting conditions.

**Table S3. Ruggedness of fjords on a grooved pin**

| Value names | | Fjord 1 (red line) | Fjord 2 (green line) | Fjord 3 (blue line) | Whole section |
|---|---|---|---|---|---|
| Curve length/μm | $f_{SL}$ | 2.275 | 1.869 | 3.162 | 20.97 |
| Straight length/μm | $f_{LV}$ | 2.037 | 1.645 | 2.745 | 19.04 |
| Fjord ruggedness | $\eta$ | 1.117 | 1.136 | 1.152 | 1.102 |
| Epoxy glue | Young contact angle /degree | $\phi_Y$ | 18.6 ±1.8 | | |
| | Minimal ruggedness required for wetting | $\eta_{min} = \dfrac{1}{\cos\phi_Y}$ | 1.058 ±0.011 | | |
| UV glue | Young contact angle /degree | $\phi_Y$ | 6.0 ±2.3 | | |
| | Minimal ruggedness required for wetting | $\eta_{min} = \dfrac{1}{\cos\phi_Y}$ | 1.006 ±0.004 | | |

## Wetting on grooved surface

The static wetting states of liquid on the conical pin were observed by fluorescent images by LSCM in Fig. 3d,f. In addition, the dynamic wetting process was observed on the grooved surface of a tungsten block. Directly observing the dynamic wetting process on the conical pin is extremely difficult because of two reasons: the nano grooves exceeded the resolution of ordinary optical microscopy, and the pin surface was not flat, thereby preventing clear focusing. Thus, we observed the dynamic wetting process on a flat tungsten surface with grooves in the width of several micrometers. The tungsten block was the same as that used in Fig. S5, whose surface was mechanically polished and had obvious textures. The topography of the tungsten surface was further measured by AFM imaging, as shown in Fig. S7b,c. A droplet with a size of 1–2 mm was placed on the tungsten surface, and the dynamic wetting process was observed by an optical microscope in real time. As shown in Fig. S7d–h, a liquid line advanced in a groove as time progressed. The liquid line had advanced 18.1 μm in 120 s, and thus it had an average speed of 0.15 μm/s.

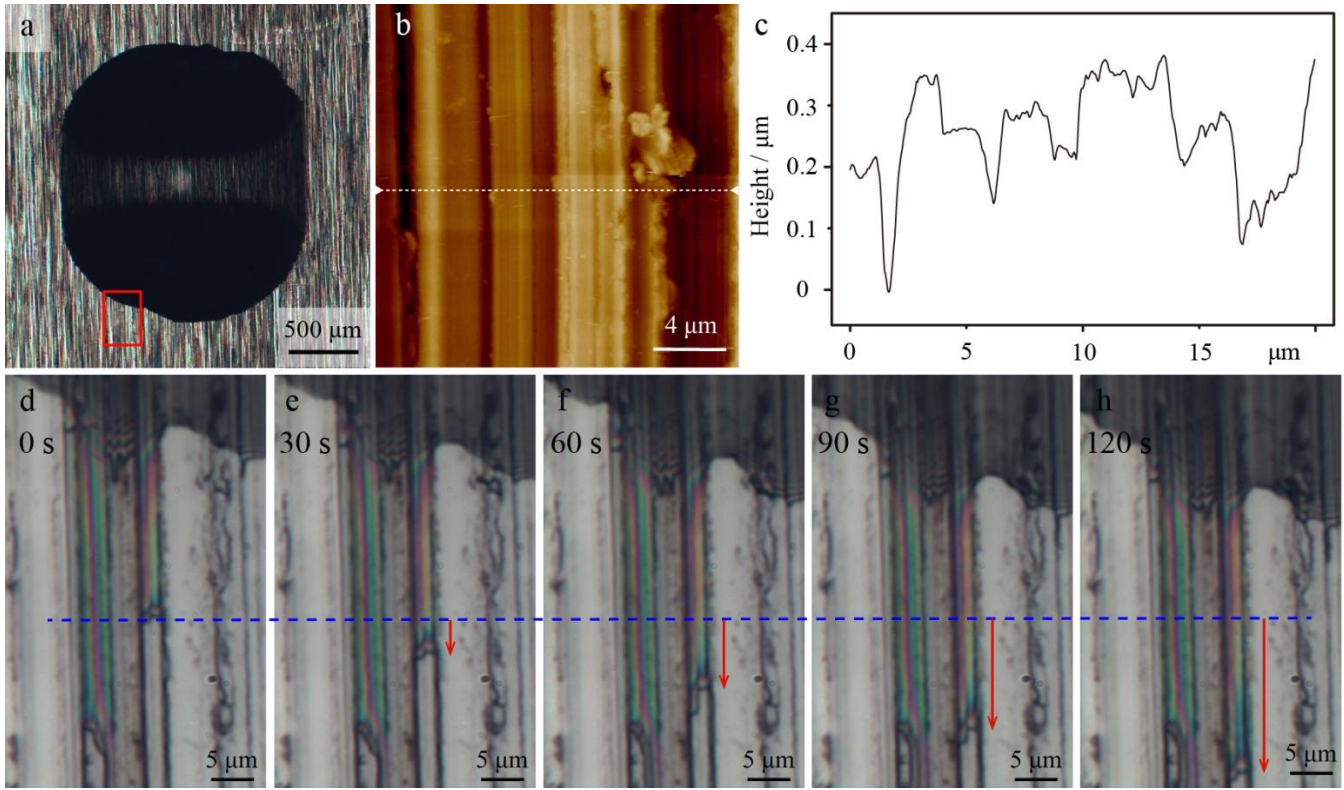

**Figure S7. Wetting state of liquid on grooved surface. a**, Epoxy glue droplet on the textured surface of a tungsten block. **b**, Typical topography of an area on the textured surface of the tungsten block imaged by AFM. **c**, Cross section in **b** at the position marked by a white dashed line. **d–h**, Images taken in time sequence of a liquid line advancing in a groove on the tungsten surface to form a fjord at 0, 30, 60, 90, and 120 s. The red rectangle in **a** indicates the position where **d–h** were obtained but has a much larger actual size than **d–h**.

## Dispensing liquid in millimeter scale

The liquid dispensing mechanism by the curvi-propulsion force to store liquid and by the fjords to transport liquid has been demonstrated in both microscale and nanoscale. Can this mechanism help liquid dispensing in larger scales such as in millimeter scale, similar to a pen writing on paper? Interestingly, writing with ink has been a basic technology from the early stage of human civilization. For example, the papyrus was used as the writing substrate in ancient Egypt about 5,000 years ago, and the brush pen was used as the writing tool in China about 3,200 years ago[33]. To investigate the phenomenon in millimeter scale, we designed two kinds of conical pins: one with smooth surface and the other with grooves, as shown in Fig. S8a. Then the pins were machined by a computer numerically controlled (CNC) turning-milling center. Each groove on the pin surface had a width of ~0.5 mm and a depth of ~0.25 mm. Considering that the capillary length of water is 2.7 mm and the machined pins had a radius of 1 mm at the end, liquid surface tension force still played an important role in the liquid dispensing of these machined pins. Black ink, which was commercially available for fountain pens, was dripped on the machined pins. After making contact with a piece of white paper, a machined pin with grooves dispensed an ink dot successfully. By contrast, the smooth pin left nothing on the white paper. This experiment demonstrated that the same mechanism also works in millimeter scale. So far, we have validated that the mechanism works in multi-scales from nanoscale, microscale, to millimeter scale.

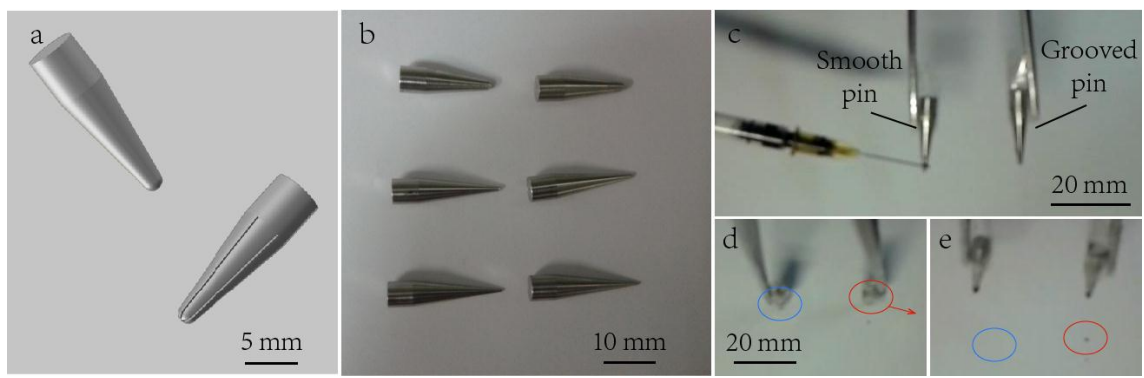

**Figure S8. Dispensing ink in millimeter scale by a machined grooved pin. a**, 3D models of two conical pins: one with a smooth surface and the other with grooves. **b**, Machined pins, three with a smooth surface and three with grooves. **c**, Fountain pen ink was dripped manually to the two pins by a syringe. **d**, The two pins were contacted on a white paper to dispense ink. **e**, After removing the two pins, the smooth pin left nothing (marked by the blue circle), whereas the grooved pin left an ink dot on the white paper (marked by the red circle). The scale bar in **e** was the same as that in **d**. The pins were made of stainless steel and machined by a CNC turning-milling center. Each grooved pin had 6 grooves circumferentially distributed on its conical surface, and each groove had a width of ~0.5 mm and a depth of ~0.25 mm.

## Image acquisition and processing

**Optical microscopes:** Three optical microscopes were used to obtain optical images. Motic PSM-1000 was used for Figs 2a, S1, S3, S5a,d, and S7a,d–h. This microscope was equipped with objectives of 2, 5, 10, 20, 50, and 100× magnification; a CMOS camera of Artray-300MI-C; and a 150 W halogen light source. Hirox KH-3000, equipped with objectives of 35 and 70× magnification, was used for Fig. S4. Olympus BX51, equipped with objectives of 4, 20, and 50× magnification, was used for Fig. S2. The optical images of Fig. S5b,c,e,f were taken by the optical microscope embedded in the contact angle meter (Data Physics OCA20).

**SEM:** Quanta FEG 450 from FEI Company was used for Figs 1d–f and 3c,e. With regard to sample preparation for SEM observation, the tungsten wire used to fabricate pins was ultrasonically washed in 1 mol/L KOH aqueous solution beforehand to remove the oxide layer and to ensure good conductivity.

The fabricated pins were carefully bonded to an aluminum bar by conductive tapes. Notably, the sharp-end parts of the pins were suspended out of the aluminum bar to avoid damage. The imaging parameters were set with an acceleration voltage of 4 kV and a spot size of 3 nm.

**LSCM:** Zeiss 710, which is equipped with a $40\times$ objective and a 543 nm He-Ne laser, was used for to obtain the fluorescent images shown in Fig. 3d,f. Each image was combined by a bright-field image with a fluorescent image. The bright-field images were used to observe the tungsten pins, because the pins did not produce any fluorescence. The bright-field images were illuminated by a white light source. The fluorescent images were used to observe the liquid distributed on the pin. The liquid was mixed with rhodamine B, thus fluorescence was produced under activation by laser. Given that the fluorescence was usually very weak, the white illumination must be turned off when taking fluorescent images. To observe the pin and the liquid simultaneously in a single image, combination processing was applied to the images. First, the initial fluorescent image was placed in the red channel (a color image has 3 channels: red, green and blue). Second, the bright-field image was initially placed in green channel, and then converted into a gray image that will occupy all three channels. Finally, the two images were arithmetically added in pixels.

**AFM:** NT-MDT NTEGRA Prima, which is equipped with scanning stage of 100 μm × 100 μm × 10 μm, was used for Figs 1g, 2b,c, S6a,b, and S7b. Contact mode was usually more reliable for topography imaging. Thus this mode was used to measure the hard tungsten block in Fig. S7b. However, using contact mode is difficult to measure the soft pin and has the risk of damaging the solidified nano glue samples. Thus, semi-contact mode was used for Figs 1g, 2b,c, and S6a,b to reduce the applied force to the samples as much as possible. The tip used for contact mode was FMG01 with a tip radius of 10 nm, a cone angle at the apex of 7 °–10 °, a force constant of 1.2–6.4 N/m, and a reflective side of Au. The tip used for semi-contact mode was HA_NC with a tip radius of <10 nm, a cone angle at the apex of ≤30 °, a force constant of 3.5 N/m, a resonant frequency of 140 kHz, and a reflective side of Au. All AFM images were initially gray images and were false colored by postprocessing. The images of glue pattern on substrate as shown in Fig. 2b,c were further processed by highlighting the glue area in green color to distinguish it from the brown substrate for clarity.

## Comparable technologies

The essential difference between DPN and our NGP technology has been discussed in the main text. A few additional details are provided in this part. The DPN was first proposed in 1999 by Piner et al.[22]

Its mechanism was further studied in 2002 to reveal the influence of temperature, humidity, and holding time[34]. The key role in DPN is the water meniscus between the AFM tip and the substrate. First, the tip was coated with a water-soluble material layer. When the tip was approaching the substrate at a distance less than 2 nm, water would be condensed from the atmosphere humidity (or collected from the residual water on substrate) at the contact point. Molecules would then dissolve from the coating into the water meniscus and diffuse to the substrate. After holding on for certain time, a few molecules would be deposited onto the substrate. The first essential difference is that the DPN uses molecule dissolving and diffusing, whereas our NGP technology uses wetting in nano grooves to transport liquid. The second essential difference is that the DPN stores material through solid coatings on the tip, whereas our NGP technology stores liquid by surface tension force on a conical pin. Thus, the two technologies have essentially different mechanisms. The DPN has advantages in depositing molecules of monolayer or a few layers. However, NGP can directly transport liquid-state materials and can work without requirements to the atmospheric conditions. Besides, we have demonstrated that NGP can work in multi-scales from nanoscale up to microscale until millimeter scale. Hence, NGP can dispense a much larger volume than a few molecules at one time. In this sense, NGP has good application potential in 3D printing[35].

We presented the developing roadmaps of major liquid dispensing technologies in Fig. 4. More details are provided in this part. Over the past several decades, liquid dispensing technologies have been evolving towards smaller, faster, simpler and applicable for more kinds of materials. Previously, the main track for liquid dispensing technology development is based on tubes or nozzles. To extrude a droplet out from a tube or nozzle, the inner pressure must be much higher than the pressure outside the tube or nozzle. According to the formula $\Delta p = 4\gamma / d$, the pressure difference is inversely proportional to the inner diameter $d$. In this formula, $\gamma$ is the coefficient of surface tension of liquid. Various methods have been used to generate the needed pressure difference. The most easily conceived solution is that an air pump is combined with a valve driving liquid to flow in a tube. However, when the diameter of the tube or nozzle is smaller than ~100 μm, excessively high air pressure will lead to significantly complex equipment and difficulty in fabricating tubes with a small diameter. Ink jet uses heat or vibration to generate the needed driving force[36], and thus, ink jet can produce droplets that are much smaller than the tube diameter without air pumps. In this way, droplets with a diameter as small as ~20 μm[9] can be dispensed successfully. Hence, ink jet has been widely used in commercial printers. EHD jet, which dispenses liquid with the assistance of an electric field, was first studied by Taylor[26]. The voltage needed to draw liquid out of a tube or nozzle can be as high as several hundred or even more than one thousand volts[37]. In 2007, Park et al. made significant progress in EHD jet achieving sub-micro resolution[9]. In 2012, Galliker et al. further reduced the resolution to 50 nm by EHD jet of liquid with nanoparticles[13]. However, the complexity increases dramatically as the resolution is improved. The primary problem is fabrication of tubes or nozzles with very small diameters. For example, the nozzles used by Galliker et al. had diameters that ranged from 0.6–1.2 μm which were nearly the technology limitation of fabrication. The secondary problem is the complex equipment needed to produce higher pressure or vibration for ink jet or to control electric circuits for EHD jet. The tertiary problem is clogging, which was also mentioned by Galliker et al. In summary, the dispensed droplets based on tubes or nozzles are downscaled at the expense of being remarkably more complex.

Dispensing liquid by AFM hollow tips can also reach a small scale of tens of nanometers. The key is the extremely small tube or nozzle that has nano hollows in an AFM tip. To the best of our knowledge, three kinds of technologies have used AFM hollow tips. The first is NADIS, which uses capillary force and surface tension force to transfer liquid without any external exerted energy[24,28,38]. The second is dispensing liquid through an AFM hollow tip with an exerted external electrical field[12]. The third is nano fountain pen[39-41], which combines the DPN with an AFM hollow tip. The third method basically utilizes the mechanism of DPN. Accordingly, this mechanism was excluded from the scope of "NADIS by AFM hollow tips" described in Fig. 4. The strategies of AFM hollow tips are even more complex than those of EHD jet because the tips are made by a procedure of extremely complicated and costly focus ion beam (FIB).

Microcontact printing (μCP)[42,43] is also widely used to transfer liquid materials or print designed patterns. The basic mechanism of μCP is similar to that of pin transfer[44] when used to transfer liquid. By further downscaled, transfer printing[45,46] could also reach micro and nano scales. These technologies were not individually placed in the roadmaps.

Specifically, pin transfer technology can be divided in two typologies[6]: solid pin transfer and split pin transfer[44]. The split pin transfer improves the continuous transfer capability because a split in the pin end can store more liquid than a solid pin. Interestingly, Chinese brushes, which were the common writing tools in ancient China, were studied by some researchers[47], and they presented that the mechanism of liquid storage and transfer in Chinese brushes was similar to that of split pin transfer. However, the basic mechanism of split pin transfer remains similar as solid pin transfer, and dispensing small droplets (<50 μm) remain difficult for split pin transfer. Therefore, we present in Fig. 4 only "pin transfer" to represent these variants of pin transfer.

We believe the best liquid dispensing technology in the future may be a combination of the two methods, EHD jet and NGP technology. Both EHD jet and the NGP technology have advantages. The speed of EHD jet by microscale tubes or nozzles is quite fast in the range of kHz. When EHD jet is further downscaled by AFM hollow tips, flow resistance increases dramatically in nanoscale; as a result, the time taken for dispensing a droplet is as long as several seconds. This speed is similarly slow as that of NGP technology. However, it is worth considering that DPN tip arrays (up to 2,500 dip-pens per square millimeter) were fabricated such that the overall DPN speed can be amplified for three or even higher orders[29,30,48]. Therefore, NGP technology has a similar potential to amplify its overall speed for orders by the parallel strategy. Moreover, the troublesome problem of clogging in tubes or nozzles rarely occurs on an NGP because small particles can barely block the numerous transport routes. On the contrary, small particles will increase the roughness of the pin surface and form new transport routes. As a result, NGP technology has an additional advantage of anti-clogging. Moreover, it is much easier to wash the outside surface of a pin than the inner surface of a tube or nozzle. In another aspect, EHD jet is also being simplified by scientists' great efforts such as the very novel pyroelectrodynamic shooting invented by Ferraro et al.[14] In terms of material applicability, EHD jet requires certain electrical conditions[49], whereas NGP technology requires certain wetting conditions. Therefore, the combined technology may need either electrical or wetting conditions.